\documentstyle[12pt,psfig]{article}
\newcommand{\Eq}[1]{Eq.\,(\ref{#1})}
\def\beq{\begin{equation}}
\def\eeq{\end{equation}}
\def\bear{\begin{eqnarray}}
\def\ear{\end{eqnarray}}

\newcommand{\T}{\theta}
\newcommand{\ahf}{\overline{^4He}}
\newcommand{\aht}{\overline{^3He}}

\newcommand{\Ref}[1]{Ref.\,\cite{#1}}
\newcommand{\Refs}[1]{Refs.\,\cite{#1}}

\author{{\large K.M.~Belotsky,}{\small $^{a,b}$}
{\large ~Yu.A.~Golubkov,}{\small $^{a,d}$}\thanks{e-mail: 
golubkov@elma01.npi.msu.su}
{\large ~M.Yu.Khlopov,}{\small $^{a,b,c}$}\thanks{e-mail: 
mkhlopov@orc.ru}
{\large ~R.V.~Konoplich,}{\small $^{a,c}$}\thanks{e-mail: 
konoplic@orc.ru}\and
{\large S.G.~Rubin}{\small $^{a,c}$}\thanks{e-mail: 
sgrubin@orc.ru}
{\large~and~A.S.Sakharov}{\small $^{a,c}$}\thanks{e-mail: 
sakhas@landau.ac.ru}}

\title{{\huge{\bf AMS-Shuttle test for antimatter stars in our 
Galaxy}}
\footnote{Invited talk on the 3d Int. Conf. on Cosmoparticle 
Physics "COSMION-97". Russia, Moscow, December 8--14, 1997.}
}
\date{{\small{\it $^a$Center for CosmoParticle Physics 
"Cosmion",
Miusskaya Pl., 4, 125047 Moscow, Russia\\ 
$^b$Institute of Applied Mathematics,
Miusskaya Pl., 4, 125047 Moscow, Russia\\
$^c$Moscow State Engineering Physics Institute 
(Technical University),  Kashirskoe Sh., 31, 115409 Moscow, 
Russia\\
$^d$Moscow State University, Institute of Nuclear Physics, 
Vorobjevy Gory,
119899 Moscow, Russia}}}

\begin{document}

\maketitle

\begin{abstract}
The AMS--Shuttle experiment is shown to be sensitive to test
the hypothesis on the existence of antimatter globular cluster 
in our Galaxy. The hypothesis follows from the analysis of possible 
tests for the mechanisms of baryosynthesis and uses antimatter 
domains in the matter dominated Universe as the probe for the physics 
underlying the origin of the matter. The total mass for the antimatter 
objects in our Galaxy is fixed from the below by the condition 
of antimatter domain survival in the matter dominated Universe 
and from above by the observed gamma ray flux. For this mass interval 
the expected fluxes of antinuclei can lead to up to ten antihelium events 
in the AMS-Shuttle experiment.
\end{abstract}

The modern big bang theory is based on inflationary models 
with baryosynthesis and nonbaryonic 
dark matter. The physical basis for all the three phenomena 
lies outside the experimentally proven theory of elementary 
particles. 
This basis follows from the extensions of the standard model.
Particle theory considers such extensions as aesthetical 
appealing
(grand unification), necessary to remove internal 
inconsistencies
in the standard model (supersymmetry, axion) or 
simply theoretically possible
(neutrino mass, lepton and baryon number violation).
Most of these theoretical ideas can not be tested directly
and particle theory considers cosmological relevance as the
important component of their indirect test.
In the absence of direct methods of study one should analyse 
the set of indirect effects, 
which specify the models of particles and cosmology. 
AMS experiment \Ref{AMS} turns to be important tool in such 
analysis.
The expected progress in the measurement of cosmic rays fluxes 
and gamma background and in the search for antinuclei and exotic 
charged
particles make this experiment important source of
information on the possible cosmological effects of particle 
theory. 
Its operation on Alpha Station will shed light on WIMP 
annihilation
in the Galaxy, on primordial black hole evaporation, on possible 
existence
of exotic charged particles and many other important clues
on the hidden parameters of the modern cosmology, following from 
the hidden sector of particle theory. The first step in this 
direction
may be done on the base of AMS-Shuttle experiment. 
	
The COSMION-ETHZ programme assumes
joint systematic study of AMS experiment as the basement for 
experimental cosmoparticle physics. 
The specifics of AMS--Shuttle experimental programme 
puts stringent restriction on the possible choice of 
cosmic signatures for the new physics. At this stage
no clear detection of positrons, gamma rays or multi GeV
antiprotons will be possible. It
makes us to reduce the analysis to 
the antinuclear signal as the profound signature of new physics
and cosmology.

The generally accepted motivation for baryon asymmetric 
Universe is the observed absence of the macroscopic amounts of 
antimatter 
up to the scales of clusters of galaxies. According to the Big 
Bang theory
baryon symmetric homogeneous mixture of matter and antimatter 
can not survive after 
local annihilation, taking place at the first millisecond of 
cosmological evolution. Spatial separation of matter and 
antimatter 
can provide their survival in the baryon symmetric Universe
but should satisfy severe constraints on the effects of 
annihilation
at the border of domains. The most recent analysis finds that 
the
size of domains should be only few times smaller than the modern 
cosmological horizon to escape the contradictions with the 
observed
gamma ray background \Ref{A}. 
In  baryon asymmetric Universe the big bang theory predicts the 
exponentially small fraction of primordial
antimatter and practically excludes the existence of primordial
antinuclei. The secondary antiprotons may appear as a result 
of cosmic ray interaction with the matter, when galaxies are 
formed.
In such interaction it is impossible to produce 
any sizeable amount of secondary antinuclei. 
Thus non exponentially small amount of antiprotons in the 
Universe 
in the period 
from $10^{-3}$ s to $10^{16}$ s and antinuclei in the modern 
Universe
are the profound signature for new phenomena, related to 
the cosmological consequences of particle theory.
The inhomogeneity of baryon excess generation 
and antibaryon excess generation as the reflection of 
this inhomogeneity represents one of the most important example 
of such 
consequences.
It turned out \Refs{1,2,3,dolg}, that practically all the existing 
mechanisms of baryogenesis can lead to 
generation of antibaryon excess in some places, the baryon 
excess, 
averaged over the whole space, being positive.  So domains of 
antimatter in baryon asymmetric Universe provide a probe for the 
physical mechanism of the matter generation.

The original Sakharov's scenario of baryosynthesis \Ref{4}
has found physical grounds in GUT models. It assumes CP 
violating effects
in 
out--of--equilibrium B--non--conserving processes, which 
generate baryon excess
proportional to CP 
violating phase. If sign and magnitude of this phase varies  in 
space, the same out--of--equilibrium B--non--conserving 
processes, 
leading to baryon asymmetry, result in $B<0$ in the regions, 
where the 
phase is negative. The same argument is appropriate for the 
models of 
baryosynthesis, based on electroweak baryon charge 
nonconservation at high temperatures as well as 
on its combination with lepton number violation processes,
related to the physics of Majorana mass of neutrino. In all 
these
approaches to baryogenesis independent on the physical nature of 
B-
nonconservation 
the inhomogeneity of baryon excess and generation of 
antibaryon excess is determined by the spatial dependence of CP 
violating phase. 

Spatial dependence of this phase is predicted in 
models of spontaneous CP violation, modified to escape the 
supermassive domain wall problem (see \Refs{1,2,dolg} and Refs. 
therein).

In this type of models CP violating phase acquires discrete 
values 
$\phi_{+}=\phi_{0}+\phi_{sp}$ and $\phi_{-}=\phi_{0}-\phi_{sp}$, 
where $\phi_{0}$ and $\phi_{sp}$ are, respectively, constant and 
spontaneously broken CP phase, and antibaryon domains appear in 
the 
regions with $\phi_{-}<0$, provided that $\phi_{sp}>\phi_{0}$.  

In models, where CP violating phase is associated with the 
amplitude of invisible axion field, spatially--variable phase 
changes 
$\phi_{vr}$ continuously from $-\pi$ to $+\pi$. The amplitude of 
axion field plays the role of $\phi_{vr}$ in the period starting 
from 
Peccei--Quinn symmetry breaking phase transition until the axion 
mass 
is switched on at $T\approx 1$ GeV. The net phase changes 
continuously and if 
baryosynthesis takes place in the considered period axion 
induced 
baryosynthesis implies continuous spatial variation of the 
baryon 
excess given by \Ref{K} 

\beq
\label{A}
b(x)=A+b\sin(\T (x))\,.
\eeq

Here $A$ is the baryon excess induced by constant CP-violating 
phase, 
which provides the global baryon asymmetry of the Universe and 
$b$ is 
the measure of axion induced asymmetry. If $b>A$, antibaryon 
excess 
is generated along the direction $\T=3\pi /2$. The stronger is 
the 
inequality $b>A$, the larger interval of $\T$ around the 
semisurface 
$\T=3\pi /2$ provides generation of antibaryon excess \Ref{K}. 
In the case
$b-A=\delta\ll A$ the antibaryon excess is proportional to 
$\delta^2$ and the relative volume occupied by it is 
proportional 
to $\delta$.  

The axion induced antibaryon excess forms the Brownian structure 
looking like an infinite 
ribbon along the infinite axion string (see \Ref{kss}). The 
minimal width of
the 
ribbon 
is of the 
order of horizon in the period of baryosynthesis and is equal to 
$m_{Pl}/T^{2}_{BS}$ at 
$T\approx T_{BS}$. At $T<T_{BS}$ this size experiences red shift 
and 
is equal to 
\beq
\label{S}
l_h(T)\approx\frac{m_{Pl}}{T_{BS}T}
\eeq
This structure is smoothed by the annihilation at the border of 
matter and antimatter domains. When the antibaryon diffusion 
scale 
exceeds $l_h(T)$ the infinite structure decays on separated 
domains. 

The size and amount of antimatter in domains, generated as the 
result
of local baryon--non--conserving out--of--equilibrium processes, 
is related to the parameters of models of CP violation and/or 
invisible axion (see \Refs{1,3,stek}).
SUSY GUT motivated mechanisms of baryon asymmetry imply flatness 
of 
superpotential relative to existence of squark condensate. Such 
a 
condensate, being formed with $B\,>\,0$, induces baryon 
asymmetry, after 
squarks decay on quarks and gluinos.  The mechanism doesn't 
fix the value and sign of B in the condensate, opening the 
possibilities for inhomogeneous baryon charge distribution and 
antibaryon domains \Refs{3,dolg}. The size and amount of antimatter in
such domains is determined by the initial distribution of 
squark condensate.

Thus the antimatter domains in the baryon asymmetric Universe 
are related to 
practically all the mechanisms of baryosynthesis, and serve 
as the probe for the mechanisms of CP violation and primordial 
baryon charge inhomogeneity. The size of domains depends on the 
parameters of these mechanisms \Refs{1,2,stek}.

General parameters of the averaged effect of the domain 
structure are the relative amount of antimatter 
$\omega_{a}=\rho_{a}/\rho_{crit}$, where $\rho_{a}$ 
is the averaged over domain density of antimatter 
and $\rho_{crit}$ is the critical density, and 
the mean size of domains (the 
characteristic scale in their distribution on sizes) or for 
small 
domain sizes the time scale of their annihilation with the 
matter.

To compare the effect of antimatter domain annihilation with the 
observational 
data one should introduce the relative amount of annihilated 
antimatter relative to the total amount of matter. One may 
easily find (see
for 
details \Ref{7}) 
that this ratio $r$ is given by:
\beq
\label{d}
r\approx\frac{bf(l\le l_a)}{A}\,,
\eeq
where $l_a$ is the maximal size of domains annihilated by the 
considered period 
and $f(l)$ is the volume fraction of domains with the size $l$.
In the case of discrete spontaneous CP violation discussed
above $b=A$. 
 
One of the features expected for antimatter domains in baryon 
asymmetrical Universe is the possibility
of diffused antiworld. It corresponds to the antibaryon 
matter density much 
smaller, than the baryon matter density.
One of the interesting consequences of diffused 
antiworld hypothesis is the possibility of unusual light 
antinuclei 
abundance. At antibaryon densities, much smaller, than the 
baryon 
density, anti--deuterium and anti--helium--3 may be more 
abundant, than 
antihelium--4. However diffused antiworld with very low 
antibaryon
density can not lead to formation of antimatter objects and
gamma ray search for annihilation in diffused antimatter clouds
is the most promising in this case. The possibility of 
antibaryon 
density in domains comparable or even higher than the
mean baryon density is much more interesting for AMS-Shuttle
programme in cosmoparticle physics.

As it was recently shown \Ref{K}, in the case when axion induced 
CP violation dominates in the process of 
Baryosynthesis, the antimatter density 
within the surviving domains should be
larger than the mean baryon density. On the other hand
the SUSY GUT squark condensate may induce large scale modulation
of this distribution. Since both axion and SUSY are considered 
as
the necessary extensions of the standard model one should 
consider
at least the combination of axion-- and squark--condesate-- 
induced
inhomogeneous baryosynthesis as the minimally realistic case. 
With the account for the other possible mechanisms for 
inhomogeneous baryosynthesis, predicted on the base of
various and generally independent extensions of the standard
model, the general analysis of possible domain distributions
is rather complicated.
Fortunately, the test for the possibility of the existence of 
antistars in
our 
Galaxy, offered in \Ref{K}, turns to be practically model 
independent and
as we 
show here may be accessible for AMS-Shuttle
Experiment.
Let us assume some distribution of antimatter domains,
which satisfies the constraints on antimatter annihilation
in the early Universe. Domains, surviving 
after such annihilation, should have 
the mass exceeding
\beq
\label{m}
M_{min}\approx (b/A)\rho_{b}l^3_a\,,
\eeq
where $\rho_b$ is the mean cosmological baryon density. The mass 
fraction
$f$ of 
such domains relative to total baryon mass is strongly model 
dependent.
Note that since the diffusion to the border of antimatter domain 
is determined on RD 
stage by the radiation friction, the surviving scale fixes the 
size of the 
surviving domain. On the other hand the constraints on the 
effects of 
annihilation put the upper limit on the mass of annihilated 
antimatter.

The modern antimatter domain distribution should be 
cut at masses given by 
the \Eq{m} due to annihilation of smaller domains and it is the 
general feature of any model of antibaryosynthesis in baryon 
asymmetrical Universe. The specific form of the domain 
distribution 
is model dependent. At the scales smaller than those given by 
\Eq{m} the 
spectrum should satisfy the constraints on the relative amount 
of 
annihilating antimatter. Provided these constraints are 
satisfied, one may consider the conditions for antimatter 
objects 
formation. One should take into account that the estimation of 
the 
annihilation scale after recombination (see \Ref{7})  gives for 
this 
scale the value close the Jeans mass in the neutral baryon gas 
after 
recombination. So the development of gravitational instability 
may 
take place in antimatter domains resulting in the formation of 
astronomical objects of antimatter.

Formation of antimatter object has the time scale being of the 
order 
of $t_f\approx (\pi G\rho)^{-1/2}$. The object is formed, 
provided 
that this time scale is smaller than the time scale of its 
collision 
with the matter clouds. The latter is the smallest in the 
beginning 
of the object formation, when the clouds forming objects have 
large 
size. 

Note that the isolated domain can not form astronomical object 
smaller than globular cluster \Ref{K}. The isolated anti--star 
can not be 
formed in matter surrounding since its formation implies the 
development of thermal instability, during which cold clouds are 
pressed by hot gas. Pressure of the hot matter gas on the 
antimatter 
cloud is accompanied by the annihilation of antimatter. 
Thus anti--stars can be 
formed in the antimatter surrounding only, what may take place 
when such 
surrounding has at least the scale of globular cluster.

One should expect to find antimatter objects among 
the oldest population of the 
Galaxy \Ref{K}. It should be in the halo, since owing to strong
annihilation of 
antimatter and matter gas the formation of secondary antimatter 
objects in
the disk component of our Galaxy is impossible. So in the estimation 
of
antimatter 
effects we can use the data on the spherical component of our 
Galaxy as
well as 
the analogy with the properties of the old population stars in 
globular
clusters 
and elliptical galaxies.

In the spherical component of our Galaxy the antimatter globular 
cluster
should 
move with high velocity (what follows from the velocity 
dispersion in halo, 
$v\approx 150$ km/s) through the matter gas with very low number 
density 
($n\approx 3\cdot 10^{-4}cm^{-3}$). Owing to small density of
antimatter gas effects of annihilation with the matter gas 
within 
the antimatter globular cluster are small. These effects, 
however, 
deserve special analysis for future search for antimatter 
cluster 
as the gamma source. 

The integral effects of antimatter cluster may be estimated by 
the 
analysis of antimatter pollution of the Galaxy by the globular 
cluster of antistars.

There are two main sources of such pollution: the antistellar 
wind 
(the mass flow from antistars) and the antimatter Supernova 
explosions. The first source provides the stationary in-flow of 
antimatter particles with the velocity $10^7\div 10^8\ cm/s$ to 
the 
Galaxy. From the analogy with the elliptical galaxies, for which 
one 
has the mass loss $10^{-12}M_{\odot }$ per Solar mass per year, 
one can 
estimate the stationary admixture of antimatter gas in the 
Galaxy and 
the contribution of its annihilation into the gamma ray 
background. 
The estimation strongly depends on the distribution of magnetic 
fields in the Galaxy, trapping charged antiparticles. Crude 
estimation of the gamma flux from the annihilation of this 
antimatter 
flux is compatible with the observed gamma background for the 
total 
mass of antimatter cluster less than $10^5 M_{\odot }$. This 
estimation 
puts upper limit on the total mass fraction of antimatter 
clusters in our 
Galaxy. Their 
integral effect should not contradict to the observed gamma ray 
background.

\begin{figure}[htb]                %        Fig.1. Ratios
\par
\centerline{\hbox{%
\psfig{figure=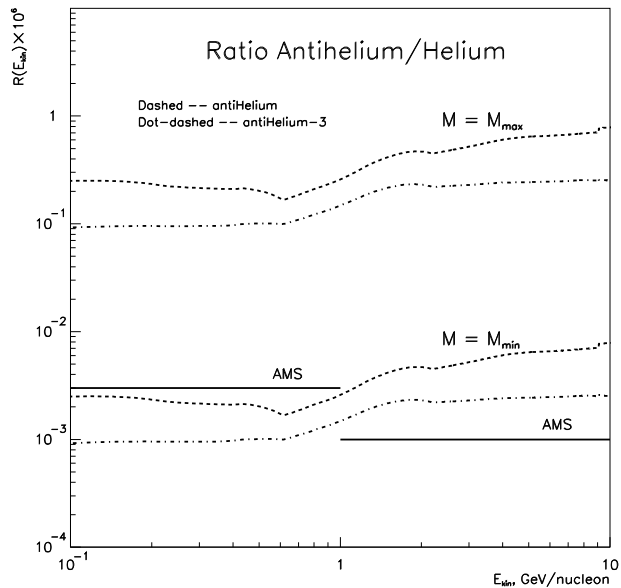,bbllx=1.5cm,bblly=5.5cm,%
bburx=19.0cm,bbury=23.0cm,clip=t,height=8.0cm}%
}}
\par
\caption{\label{ratios}
Ratios of fluxes $\ahf /^4He$ (dashed) and
$\aht /^4He$ (dot--dashed). Two upper curves correspond to the 
case of the
maximal possible mass of antimatter globular cluster 
$M_{max}\,=\,10^5\,M_{\odot }$ 
and the two lower curves to the case of the minimal possible 
mass 
of such cluster $M_{min}\,=\,10^3\,M_{\odot }$. 
The results of calculations are compared with the expected
sensitivity of AMS experiment \Ref{AMS} (solid lines).
}
\end{figure}

The uncertainty in the distribution of magnetic fields causes 
even more
problems in the reliable 
estimation of the expected flux of antinuclei in cosmic rays. It 
is 
also accomplished by the uncertainty 
in the mechanism of cosmic ray acceleration. The relative 
contribution of
disc and halo particles into 
the cosmic ray spectrum is also unknown. To get some feeling of 
the
expected effect we may assume that the 
mechanisms of acceleration of matter and antimatter cosmic rays 
are similar
and that the contribution 
of antinuclei into the cosmic ray fluxes is proportional to the 
mass ratio
of globular cluster and Galaxy. Putting together the lower limit 
on the mass
of the antimatter globular cluster from the condition of 
survival 
of antimatter domain 
and the upper limit on this mass following from the observed 
gamma ray 
background one obtains \Ref{K}  the expected flux of antihelium 
nuclei in
the cosmic rays with the energy exceeding $0.5$ GeV/nucleon to 
be 
$10^{-8}\div 10^{-6}$ of helium 
nuclei observed in the cosmic rays. The results of numerical 
calculation of
the expected antihelium flux in \Ref{7} together with the 
expected
sensitivity of AMS experiment \Ref{AMS} are given in 
Fig.\ref{ratios}.

\vspace*{0.3cm}

Such estimation assumes that annihilation does not influence the 
antinuclei
composition of cosmic rays, which fact may take place if the 
cosmic ray
antinuclei are initially relativistic. If the process of 
acceleration 
takes place outside the antimatter globular cluster one should 
take into account the Coulomb effects in the annihilation cross 
section 
of non relativistic antinuclei, which may lead to
suppression of their expected flux.

On the other side the antinuclei annihilation invokes new factor 
in the problem of their acceleration, which is evidently absent 
in the case of cosmic ray nuclei. 
This factor may play very important role in the account for 
antimatter 
Supernovae as the possible source of cosmic ray antinuclei. From 
the
analogy with elliptical galaxies one may expect \Ref{K} that 
in the antimatter globular cluster 
Supernovae of the I type should explode with the frequency about 
$2\cdot 10^{-13}/{M_{\odot }}$ per year. On the base of 
theoretical models 
and observational  data on SNI (cf. \Ref{11}) one expects in 
such 
explosion the expansion of a shell with the mass of about $1.4 
M_{\odot }$ 
and velocity distribution up to $2\cdot 10^9\ cm/s$. 
The internal layers with the velocity $v\,<\,8\cdot 10^8\ cm/s$ 
contain anti-iron $^{56}Fe$ and the outer layers with higher 
velocity
contain lighter elements such as anti-calcium or anti-silicon. 
Another important property of Supernovae of the I type is the 
absence 
of hydrogen lines in their spectra. 
Theoretically it is explained as the absence of
hydrogen mantle in Presupernova. In the case of antimatter 
Supernova
it may lead to strong relative enhancement of antinuclei 
relatively
to the antiprotons in the cosmic rays. Note that similar effect
is suppressed in the nuclear component of cosmic rays, since 
Supernovae of the II type are also related to the matter cosmic 
ray origin
in our Galaxy, in which massive hydrogen mantles (with the
mass up to few solar masses) are accelerated.  

In the contrast with the ordinary Supernova the expanding 
antimatter shell
is 
not 
decelerated owing to acquiring the interstellar matter gas and 
is not
stopped by its pressure but annihilate with it \Ref{K}. 
In the result of annihilation with 
hydrogen, of which the matter gas is dominantly composed, semi-
relativistic
antinuclei fragments are produced. The reliable analysis of such 
cascade of
antinuclei annihilation may be based on the theoretical models 
and
experimental data on antiproton nucleus interaction. 
This programme is now under way.
The important qualitative result is the possible nontrivial 
contribution into
the fluxes of cosmic ray antinuclei with $Z\le 14$ and the 
enhancement of 
antihelium flux.

Another important qualitative effect of annihilation in the 
expected
composition of cosmic ray antinuclei is the possible presence 
of significant fraction of anti--helium--3. 
One can expect \Ref{K}  this fraction to be of the order of 
$0.2$ of the expected flux of anti--helium--4. This estimation 
follows 
from the experimental data on antiproton--helium interaction 
measured 
in the experiment PS 179 at LEAR CERN \Ref{12}. 

The estimations of \Ref{K}  assumed stationary in-flow of 
antimatter in
the cosmic rays. In case Supernovae play the dominant role 
in the cosmic ray origin the in--flow is defined by their 
frequency. 
One may find from \Ref{K}  that the interval of possible masses 
of antimatter cluster $(3\cdot 10^{3}\div 10^5)\ M_{\odot }$
gives the time scale of antimatter in--flow 
$1.6\cdot 10^{9}\div 5\cdot10^{7}$ years, which value exceeds 
the generally 
estimated life time of cosmic rays in the Galaxy. 
The succession of antinuclear annihilations may result 
in this case in the dominant contribution of antihelium--3 into 
the expected antinuclear flux.

To conclude, with all the reservations mentioned above on the 
base of the 
hypothesis on antimatter globular cluster in our Galaxy one may 
predict at
the 
level of the expected 600 antiprotons up to ten antihelium 
events in the
AMS--Shuttle experiment. Their detection will be exciting 
indication favouring
this 
hypothesis. Even the upper limit on antihelium flux will lead to 
important 
constraint on the fundamental parameters of particle theory and 
cosmology
to be 
discussed in our successive publications.

\bigskip \bigskip
\noindent {\bf Acknowledgments} \par
\noindent The Russian side of COSMION--ETHZ collaboration 
expresses its 
gratitude to ETHZ for the permanent support of studies 
undertaken in the 
framework of International projects "Astrodamus" and " COSMION--
ETHZ".

\end{document}